\documentclass[a4paper,fleqn,usenatbib,useAMS,usefloat]{mnras}
\usepackage{mathptmx}
\usepackage[normalem]{ulem}
\usepackage[T1]{fontenc}
\usepackage{ae,aecompl}
\usepackage{longtable}

\usepackage{graphicx}	% Including figure files
\usepackage{epstopdf}
\usepackage{xcolor}
\usepackage{amsmath}	% Advanced maths commands
\usepackage{amssymb}	% Extra maths symbols
\usepackage{subfig}
\usepackage{ulem}
\usepackage{booktabs}
\usepackage{float}
\usepackage{array}
\usepackage{caption}
\usepackage{scalerel}
\usepackage{hyperref}
\hypersetup{
    colorlinks=true,
    linkcolor=blue,
}
\title[Modified minimum mass ratio]
{Search for Dormant Black Holes in Ellipsoidal Variables II. A Binary Modified Minimum Mass Ratio}
\author[Gomel, Faigler and Mazeh]
{Roy Gomel,\thanks{E-mail:
roygomel@tauex.tau.ac.il}
Simchon Faigler
and 
Tsevi Mazeh
\\
School of Physics and Astronomy, %Raymond and Beverly Sackler Faculty of Exact Sciences, \\
Tel Aviv University, Tel Aviv, 6997801, Israel\\
}

\date{Accepted 2021 April 8. Received 2021 March 25; in original form 2021 January 31}

\begin{document}
\label{firstpage}
\pagerange{\pageref{firstpage}--\pageref{lastpage}}
\maketitle
%
%======================================
\begin{abstract}
This is the second  of a series of papers that focuses on searching large sets of photometric light curves for evidence of close binaries with a dormant black hole,
and, in some cases, a dormant neutron star.
The detection of such a binary is based on identifying a star that displays a large ellipsoidal periodic modulation, induced by tidal interaction with its companion.
Based on the observed ellipsoidal amplitude and the primary mass and radius, one can derive a minimum mass ratio of the binary. 
%defined as the mass ratio obtained for an inclination of $90^{\circ}$.
%
A binary with a minimum mass ratio significantly larger than unity might be a candidate for having a dormant compact-object companion. 
Unfortunately, the photometric search is hampered by the fact that in many cases the primary mass and radius are not well known.
In this paper we present a simple approach that circumvents this problem by suggesting a robust {\it modified} minimum mass ratio, assuming the primary fills its Roche lobe. 
The newly defined modified minimum mass ratio is always {\it smaller} than the minimum mass ratio, which is, in its turn, smaller than the actual mass ratio. Therefore, binaries with a modified minimum mass ratio larger than unity are candidates for having a compact-object secondary.
%even though we cannot reliably constrain their primary mass and radius. 
\end{abstract}
%===================================

%

%===========================
\section{Introduction}
\label{sec:intro}
%===========================

This is the second paper of a series \citep*[Paper I:][]{gomel21} that focuses on searching large sets of stellar photometric light curves for evidence of close binaries with a dormant black-hole (BH), and, in some cases, a dormant neutron-star (NS) secondary. 
Such systems were not discovered yet as they do not emit X-rays, either because there is no mass transfer between the primary star and the compact object, or because the accretion disc is in a quiescent state.
The detection of such a dormant binary is based on identifying a star that displays a large ellipsoidal periodic modulation, induced by tidal interaction with its companion.

The tidal interaction distorts the stellar surface, resulting in an observed modulation with half the orbital period. 
The amplitude of the ellipsoidal modulation depends, for a circular orbit, on the stellar radius, the semi-major axis and the inclination of the orbit, the binary mass ratio, and, to some extent, on the stellar temperature
\citep[e.g.,][]{kopal59, morris85, bochkarev79,
morris93}. 

Based on the observed ellipsoidal amplitude and the mass and radius of the more luminous primary star, one can derive a minimum mass ratio (MMR) of the binary, defined as the mass ratio obtained for an inclination of $90^{\circ}$, provided most of the light is coming from the primary star only \citep[e.g.,][]{faigler11,faigler15}. This is similar to the MMR derived for a single-line spectroscopic binary \citep[e.g.,][]{mazeh92,boffin93,shahaf17}.

A binary with an MMR significantly larger than unity may be a candidate for having a dormant compact-object companion. 
This is so because for a binary with two main-sequence (MS) stars we expect the more massive star to overshine its companion. The fact that we observe the less massive star in the system might indicate that the other star is a compact object. 

Note, however, that this is not always the case. Algol-type binaries, which probably went through a mass-transfer phase during their evolution \citep[e.g.,][]{erdem14,dervi18,chen20}, are famous counterexamples. In these binaries, a giant or a sub-giant primary is the less massive component, but nevertheless the brighter star of the system \citep[e.g.,][]{nelson01, budding04, mennekens17, negu18}.
Indeed, the recently suggested systems consisting of an evolved primary and a dormant compact-object secondary \citep[e.g.,][]{thompson19, liu19, rivinius20, jayasinghe21}, could be Algol-type binaries \citep[e.g.,][]{van-den-Heuvel20, irrgang20, shenar20, bodensteiner20, mazeh20, el-badry20}.

Therefore, a more restricted list of candidates should include only stars that lie on or near the main sequence, with MMRs larger than unity. This applies to spectroscopic and photometric candidates alike.

Unfortunately, the photometric search for dormant BHs is hampered by the fact that in many cases the primary mass and radius are not well known, even in the Gaia \citep{Gaia2016} EDR3 era \citep[][]{gaiaEdr3}, when the stellar parallaxes are better known \citep[e.g.,][]{stassun-torres21}. In many cases, the Gaia parallaxes are still not accurate enough \citep{fabricius20}, and the constraints on the extinction are also not known to high precision \citep{green18}. This is especially true towards the Galactic bulge, where most of the known Galactic binaries that contain BHs reside \citep[e.g.,][]{BlackCAT16,tetarenko16}.
In such cases, the MMR cannot be reliably obtained, especially because the ellipsoidal modulation depends on the stellar radius to the third power. 

In this paper we present a simple approach that circumvents this problem by deriving a robust {\it modified} minimum mass ratio (mMMR), assuming the primary fills its Roche lobe. 
This is done with the correction to the \citet[][MN93]{morris93} formula, developed in  
\hyperlink{cite.gomel21}{paper I},
%the first paper of the series \citep*[][]{gomel21}, 
where
we extended the \hyperlink{cite.morris93}{MN93} formalism to include binaries with a high fillout factor, based on the PHOEBE software package \citep{prsa05,prsa16,horvat18,jones19}.

The newly defined mMMR is always {\it smaller} than the MMR, which is, in its turn, smaller than the actual mass ratio. Therefore, binaries with mMMR larger than unity are good candidates for having a compact object secondary, even though we cannot reliably constrain their primary mass and radius. 

Interestingly, one can reverse the process and use the observed ellipsoidal amplitude to impose a mass-radius relation on the primary, according to which the radius goes like the cube root of the mass.
The coefficient of this relation depends on the fillout factor and binary inclination, which are not known. However, the coefficient values of these relations are quite limited in most cases. We can use the range of allowed mass-radius relations, together with some additional constraints, to obtain some boundaries of the mass and radius of the primary. One can use, for example, an MS mass-radius relation, if we know the primary star is on or not too far from the MS. These boundaries can help us better understand the system under study.

The modified minimum mass ratio can be an efficient tool for searching large sets of stellar light curves for binaries with dormant BHs, and sometimes NSs. Short-period ellipsoidal binaries with dormant compact objects can substantially enlarge the population of known systems with BH and NS companions.

Section~\ref{sec:ModifiedMassRatio} presents our modified minimum mass ratio, Section~\ref{sec:MassRadius} explains the newly derived mass-radius relation for the primary, 
Section~\ref{sec:examples} analyses two low-mass x-ray binaries (LMXB) to illustrate the potential of our technique and Section~\ref{sec:discussion} discusses our results.

%==============================
\section{Modified Minimum Mass Ratio}
\label{sec:ModifiedMassRatio}
%==============================

Consider a binary system of two stars with masses $M_{_1}$ and $M_{_2}$, for which we observe the light coming from the primary star, $M_{_1}$, only. 
Note that according to the terminology used here the secondary, $M_{_2}$, is the assumed unseen companion.
We are interested in the primary's relative ellipsoidal modulation at a given optical band, caused by the tidal interaction with the secondary.

An approximation for the first four harmonics of the primary-star ellipsoidal modulation is given by \hyperlink{cite.morris93}{MN93}, assuming a tidally-locked primary in a circular orbit, adopting linear limb- and gravity-darkening laws.
The leading term of the approximation is
%
%----------------------------------------------------------
\begin{equation}
\frac{\Delta L}{\overline{L}} \approx
\frac{1}{\overline{L}/L_0}\alpha_\mathrm{2}
\left(\frac{R_1}{a}\right)^3 
q\, \sin^2 i\, \cos 2\phi 
\equiv A_2\, cos 2\phi
\label{eq:ellip1}
\end{equation}
%------------------------------------------------------------------
where 
$q$ is the secondary-to-primary mass ratio $q\equiv M_{_2}/M_{_1}$, 
$E(q)$ is the \cite{eggleton83} approximation for the volume-averaged Roche-lobe radius in binary semi-major axis units, 
$i$ is the orbital inclination, 
the $\alpha_2$ coefficient is a function of the linear limb- and gravity-darkening coefficients of the primary of order unity,
$\overline{L}$ is the average luminosity of the star, and $L_{_0}$ being the stellar brightness with no secondary.
The angle $\phi$ represents the orbital phase, with $\phi$=0 defined to be 
at superior conjunction, and $A_2$ is the amplitude of the ellipsoidal at half the orbital period.

In \hyperlink{cite.gomel21}{paper I}
we express the \hyperlink{cite.morris93}{MN93} approximation with the 
fillout factor $f\equiv R_{_1}/R_{\scalebox{0.6}{\rm Roche,1}}$,
where $R_{_1}$ is the primary radius and $R_{\scalebox{0.6}{\rm Roche,1}}$ is the volume-averaged radius of the primary Roche lobe.
The leading \hyperlink{cite.morris93}{MN93} term, proportional to 
$f^3$, can be written, with a correction term derived by \hyperlink{cite.gomel21}{paper I}, as
%
%---------------------------------------------------------------
\begin{equation}
A_2 \approx
\frac{1}{\overline{L}/L_{_0}}\alpha_\mathrm{2} \
f^3 E^3(q) \ q \ \sin^2 i \ C(q,f) \, ,
\label{eq:qminEq}
\end{equation}
%-------------------------------------------------------------
where the correction coefficient $C(q,f)$ starts at $1$ for $f = 0$ (no correction), as expected, and rises monotonically as $f\to 1$, obtaining a value of $\sim$ 1.5 at $f \gtrsim 0.9$. 

Equation~(\ref{eq:qminEq}) expresses the ellipsoidal amplitude with three unknowns --- the fillout factor, the mass ratio and the inclination. Consider a binary that its ellipsoidal amplitude is determined by the observations and assume for a minute that we can estimate its fillout factor too.
The fillout factor can be estimated by using the orbital period, and mass and radius of the primary, as is shown below.
In such a case, one can obtain a minimum mass ratio, 
 $q_{\rm min}$,
using Equation~(\ref{eq:qminEq}) for an inclination of $90^{\circ}$.
%one cannot derive the mass ratio. 
% we cannot use Equation~(\ref{eq:fegg}), and are left with Equation~(\ref{eq:qminEq}) alone.

However, in many cases, the stellar mass or radius of the primary cannot be reliably estimated, and therefore we do not know the fillout factor. Nevertheless, 
we can still obtain a stringent lower limit for $q$ --- defined as the {\it modified} minimum mass ratio, mMMR or $\hat{q}_{\rm min}$, by assuming the maximum values for $f$ and $\sin i$, namely that the primary star fills its Roche lobe, $f\sim 1$, and $\sin i = 1$. One can show that the resulting $\hat{q}_{\rm min}$ is always smaller than $q_{\rm min}$, attained when one assumes only one factor of Equation~(\ref{eq:qminEq}) --- $\sin i$, to have its maximal value.

\begin{figure} 
\centering
{  \includegraphics[scale=0.6]{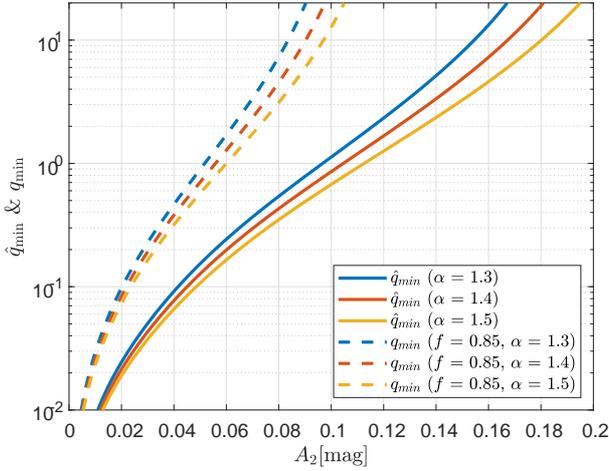}  }
\caption{
Modified minimum mass ratio, $\hat{q}_{\rm min}$, as a function of the ellipsoidal amplitude, $A_2$, assuming $f$ = 0.98 (solid line), using three typical values of $\alpha_2$ for the $V$-band. 
Note that the resulting $\hat{q}_{\rm min}$ gets larger than unity as $A_2\gtrsim$0.1, {\it independent from the orbital period and mass and radius of the primary}.
Dashed lines present the corresponding MMR, ${q}_{\rm min}$, derived for $f$ = 0.85 with the same $\alpha_2$ values.
}
\label{fig:A2_qmin}
\end{figure}

In Fig.~\ref{fig:A2_qmin} we plot $\hat{q}_{\rm min}$ as a function of $A_2$, using Equation~(\ref{eq:qminEq}) with $f\simeq 1$, and three typical $\alpha_2$ values for the $V$-band \citep{claret11}. 
We set $f$ to be $0.98$, in order to be able to use the analytical approximation of \hyperlink{cite.gomel21}{paper I}.
The resulting $\hat{q}_{\rm min}$ depends only on $A_2$, and gets larger than unity as $A_2\gtrsim$0.1,
{\it independent from the orbital period and mass and radius of the primary}. 

To display the effect of the maximal value we attributed for $f$, the figure also shows the MMR derived
from Equation~(\ref{eq:qminEq})
for $f = 0.85$, with the same $\alpha_2$ values. The diagram assumes $\sin i=1$ for all functions. 
One can notice that for, say, $A_2=0.08$, $\hat{q}_{\rm min}\sim0.4$ while $q_{\rm min}(f=0.85)\sim 5$. This emphasizes the fact that the mMMR can substantially underestimate the mass ratio, even if the primary is close to filling its Roche lobe. 

The monotonic relation between $\hat{q}_{\rm min}$ and $A_2$ imposed by Equation~(\ref{eq:qminEq}) 
suggests that a binary with ellipsoidal amplitude $A_2 \gtrsim 0.1$ probably contains an unseen companion more massive than the primary.
Furthermore, Fig.~\ref{fig:A2_qmin} suggests that such systems are expected to have a primary fillout factor not far from unity --- say, $f \gtrsim$ 0.85. 

%------------------------------------------------------------------
%Figure 1
%-----------
%------------------------------------------------------------------

%------------------------------------------------------------------
%\begin{figure} 
%	\centering
%	{  \includegraphics[scale=0.8]{A2-fminS.eps}  }
%\caption{ Minimum fillout factor as a function of the ellipsoidal semi amplitude for three typical values of $\alpha_2$ in V-band (see text), assuming $q_{\rm min}$ = 20.
%}
%\label{fig:A2_fmin}
%\end{figure}
%------------------------------------------------------------------

\newpage

%======================================================
\section{Mass-radius relation for the primary}
\label{sec:MassRadius}
%======================================================

%================================================
To obtain another constraint on the mass ratio that does use the binary period and the mass and radius of the primary one can use the \cite{eggleton83} approximation for the volume-averaged Roche-lobe radius and Kepler's third law 
%----------------------------------------------------
\begin{equation}
\left( \frac{a}{R_\odot} \right) ^3 =  4.2^3
\left( \frac{M_{_1}}{M_\odot} \right)
\left( \frac{P}{1\, \rm day} \right) ^2
\left( 1+q \right) \, ,
\label{eq:kepler}
\end{equation}
%---------------------------------------------------------
where ${M_\odot}$ and ${R_\odot}$ are the solar mass and radius,
$a$ is the semi-major axis and $P$ is the orbital period, to obtain another relation between $q$ and $f$:
%
%----------------------------------------------------
\begin{equation}
f \left( 1+q \right) ^{1/3} E(q) = 
0.238 
\left( \frac{R_{_1}}{R_\odot} \right) \,
\left( \frac{M_{_1}}{M_\odot} \right) ^{-1/3}
\left( \frac{P}{1\, \rm day} \right) ^{-2/3} \ .
\label{eq:fegg}
\end{equation}
%--------------------------------------------------

For a given set of $A_2$, $P$, $\alpha_2$, $M_{_1}$,  and $ R_{_1}$, Equations~(\ref{eq:qminEq}) and (\ref{eq:fegg}) can be solved to obtain the MMR, $q_{\rm min}$, and the corresponding $f$ of the system, assuming $\sin i = 1$. 
%=======================================================

\begin{figure*} 
	\centering
	{  \includegraphics[scale=0.8]{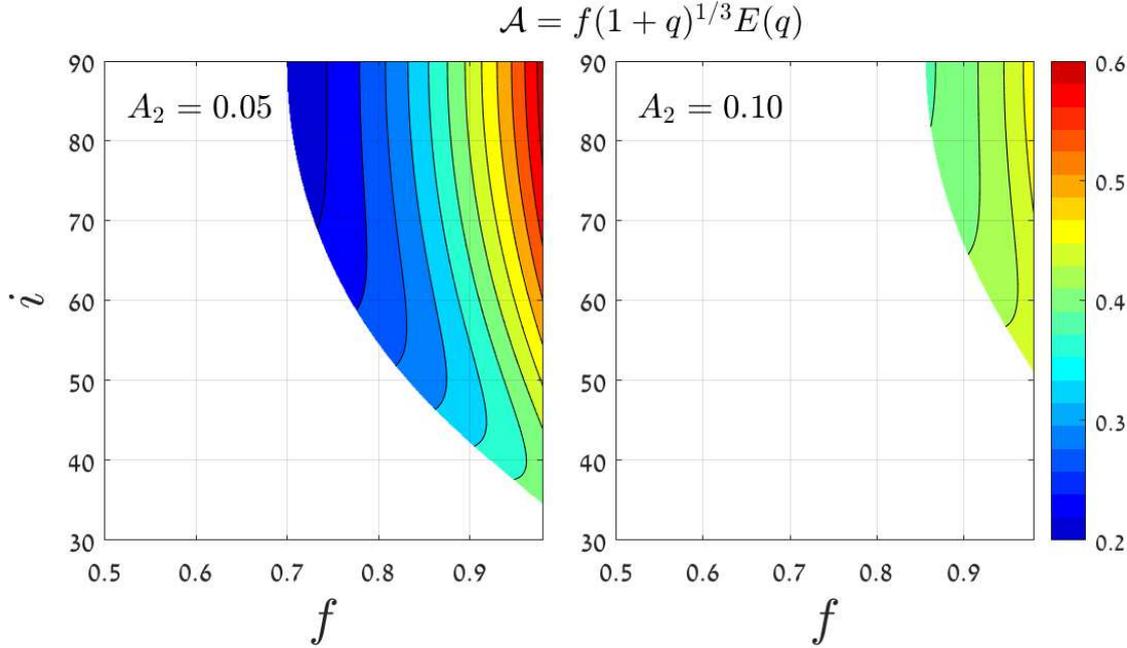}  }
\caption{Numerically derived 2-D maps of $\mathcal{A}=f \left( 1+q \right) ^{1/3} E(q)$ as a function of the Roche-lobe fillout factor of the primary, $f$, and the orbital inclination, $i$, for two values of $A_2$. It is evident that the permitted $\mathcal{A}$ range is quite limited and depends on $A_2$. The diagrams were derived using Equation~(\ref{eq:qminEq}) with $\alpha_2$ of $1.4$. Contour lines with equally-spaced $\mathcal{A}$ values are drawn from 0.2 upwards in steps of 0.02. $\mathcal{A}$ attains it maximum when $(f,i,q)$ = $(0.98,90^{\circ},0.1)$ or $(0.98,90^{\circ},0.9)$, and its minimum when $(f,i,q)$ = $(0.70,90^{\circ},20.0)$ or $(0.86,90^{\circ},19.6)$, for the left and right panel, respectively.
}
\label{fig:Factor}
\end{figure*}

In fact, 
Equation~(\ref{eq:fegg}) enables us to impose an interesting  constraint on the mass-radius relation of the primary star:
%when only the orbital period and the ellipsoidal amplitude are known:

%----------------------------------------------------
% Eq. 4
%------------
\begin{equation}
\frac{R_{_1}}{R_\odot} = 
4.2\,\left[{f}\,
 \left( 1+q \right) ^{1/3} E(q) \right]\,
\left( \frac{P}{1\, \rm day} \right) ^{2/3} 
\left( \frac{M_{_1}}{M_\odot} \right) ^{1/3}\label{eq:RvsM} \, .
\end{equation} 
%--------------------------------------------------
%
For a given ellipsoidal amplitude, which allows us to obtain $q$ for assumed values of $f$ and $i$, one can derive 
$\mathcal{A}(f,i)=f \left( 1+q \right) ^{1/3} E(q)$. 
Interestingly, the permitted $\mathcal{A}$ range is quite limited, depending on $A_2$. This is demonstrated in Fig.~\ref{fig:Factor}, where we present two examples of numerically derived 2-D maps of $\mathcal{A}$ for two values of $A_2$ --- $0.05$ and $0.1$. 
%Somewhat arbitrarily, we choose the maximum $q$ to be $20$ in order to get the minimal $\mathcal{A}$.

\begin{figure*} 
\centering
{  \includegraphics[scale=0.8]{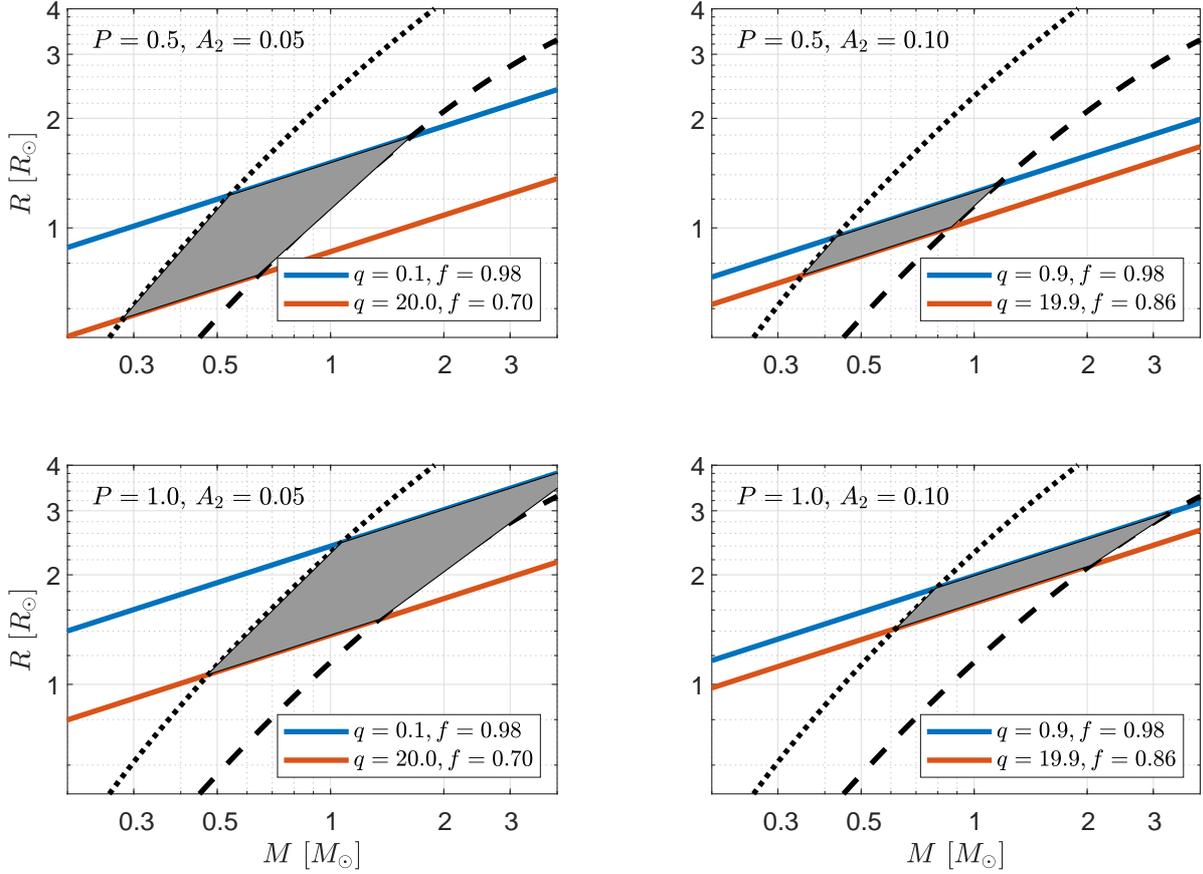}  }
\caption{
Mass-radius relations of ellipsoidal variables for different sets of ($A_2,P$). Blue and red lines correspond to the highest and lowest slopes of the mass-radius relations respectively, obtained for mass ratio and fillout factor indicated in the legend, and inclination of $90^{\circ}$. Both relations are plotted with $\alpha_2 = 1.4$. Dashed (dotted) line presents $R=R_{\scalebox{0.6}{ZAMS}}$ ($R=2R_{\scalebox{0.6}{ZAMS}}$) of \citet{eker18}, smoothed by a fifth-degree polynomial for convenience. We expect the primary star of these virtual systems to reside within the shaded region in most cases.
}
\label{fig:M_R}
\end{figure*}

The figure shows that $\mathcal{A}$ attains its maximum when $f$ and $\sin i$ are at unity, with $q \sim 0.1$ (left panel) or $\sim 0.9$ (right panel). Its minimum is obtained at $f \sim 0.7$ (left panel) or $ \sim 0.85$ (right panel), but still at $\sin i=1$ and $q \sim 20$.
%which is the upper limit of the GRID we used.
In any given binary, we can use the maximum and minimum of $\mathcal{A}$, which depend only on $A_2$, to impose two limiting constraints on the mass-radius relation of the primary. 
%To find the minimum and maximum of the left-hand side of Equation~(\ref{eq:fegg}), we use a linear 2D grid in $i$ (from $1^{\circ}$ to $90^{\circ}$ in steps of $0.1^{\circ}$) and $f$ (from $0.1$ to $0.98$ in steps of $10^{-4}$). For each grid point we calculated $q$ from Equation~(\ref{eq:qminEq}), and excluded points with no solution or $q > 20$. Then, for each of the points, we calculated the left-hand side of Equation~(\ref{eq:fegg}) and found numerically its minimum and maximum. 
These are used in 
Fig.~\ref{fig:M_R} to plot the resulting limiting mass-radius relations for two different values of $P$ and $A_2$. In the same plots we present mass-radius relations for zero-age main-sequence stars, $R=R_{\scalebox{0.6}{\rm ZAMS}}$, and for $R = 2R_{\scalebox{0.6}{\rm ZAMS}}$. We expect the primary star of the four virtual binaries to be within the region bounded by the four graphs.
%

%------------------------------------------------------------------
%Figure 2
%------------
%------------------------------------------------------------------
%

%------------------------------------------------------------------
%Figure 3
%-----------
%------------------------------------------------------------------

\newpage

%============================================
\section{Two Examples of Low-Mass X-ray Binaries}
\label{sec:examples}
%=============================================

%{\bf Most of the active low-mass black-hole binaries (LMBHBs) were not detected yet, because they were in between outbursts at the time of the X-ray surveys. In addition, LMBHBs without an extensive mass transfer, because the optical counterpart is in an earlier evolutionary phase, are X-ray quiescent and could not have been detected.}

In this section, we use two examples of known LMXBs to present the potential of our technique. Obviously, the primary star of these binaries fills its Roche lobe, $f\sim 1$, to allow mass transfer that generates accretion-driven X-rays. We are interested if these two systems could have been detected without observing the X-rays. 

To simulate our search we have to consider the photometry of these systems in the quiescent phase, when their accretion discs do not contribute to the optical brightness. 
Instead, we use the photometric analysis performed on these two systems,  which was able to separate the variability of the primary from the accretion-disc contribution. We then apply our technique to the periodic modulations, assuming they originate from the ellipsoidal effect.
%------------------------------
\subsection{A0620-00}
\label{subsec:A0620}
%------------------------------
A0620-00 is a low-mass X-ray binary consisting of a K-star primary orbiting around a compact companion, presumably a BH, with an orbital period of $P = 0.323$ day \citep{johannsen09}. A combined analysis of the spectroscopic and photometric data of the soft X-ray transient A0620-00 was performed by \citet{cantrell10}, taking into account the non-trivial BH-disk contribution to the light curve. They assumed that the primary star fills its Roche lobe and derived an orbital inclination of $i=59 \pm 0.9$, a stellar mass of $M_{_1} = 0.40 \pm 0.045 M_{\odot}$, and an estimated BH mass of $M_{_2} = 6.61 \pm 0.25 M_{\odot}$. 
According to their analysis, the $V$-band ellipsoidal component presented a semi-amplitude of about $A_2 \simeq 0.1$ mag, as illustrated in \citet[][Fig.~5]{cantrell10}.

Had the system been a short-period binary with a dormant BH component, we probably  would be able to identify it as a BH candidate based on the large amplitude of the ellipsoidal modulation, with 
$\hat{q}_{\rm min} \sim 1$.

\begin{figure} 
	\centering
	{  \includegraphics[height=5cm, width=8cm]{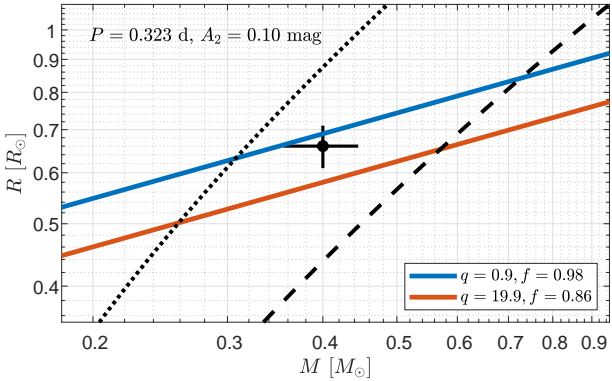}  }
	\caption{ Mass-radius relations of A0620-00, derived for $P = 0.323$ d, $A_2 = 0.1$ mag, and a typical $\alpha_2$ value of $1.4$ for the V band. The origin of each plotted line is described in the caption of Fig.~\ref{fig:M_R}. The black circle marks A0620-00 position using the values of \citet{cantrell10} and adopting a stellar-radius uncertainty of 0.05 $R_{\odot}$. 
	}
	\label{fig:A0620}
\end{figure}

Using $P$ and $A_2$ and Equations~(\ref{eq:qminEq}) \& (\ref{eq:fegg}), we derive two mass-radius relations for this system, shown in Fig.~\ref{fig:A0620}. 
The actual location of the primary is presented on the mass-radius diagram using the values of \citet{cantrell10}, within the permitted region of our analysis. The diagram suggests that the star is very close to filling its Roche lobe, and its radius is about $1.5$ larger than the corresponding ZAMS value, consistent with the results of \citet{cantrell10}.

%
%------------------------------------------------------------------
%------------------------------------------------------------------

%---------------------------------------------
\subsection{GRS 1124-683}
\label{subsec:GRS1124}
%---------------------------------------------

 Another example is GRS 1124-683, consisting of a K-star primary orbiting a compact companion, presumably a BH, with an orbital period of $P = 0.433$ day \citep{wu15}. A similar analysis of \citet{wu16} determined an orbital inclination of $i=43.2^{+2.1}_{-2.7}$, a stellar mass of $M_{_1} = 0.89^{+0.08}_{-0.11} \ M_{\odot}$ 
 and a BH mass of $M_{_2} = 11.0^{+2.1}_{-1.4} \ M_{\odot}$, assuming that the primary star fills its Roche lobe. The $I$-band ellipsoidal component presented a semi-amplitude of about $A_2 \simeq 0.07$ mag, as illustrated in \citet[][Fig.~6]{wu16}. 
 
Had the system been a short-period binary with a {\it dormant} BH component, we probably would  identify it as a low-probability candidate based on the amplitude of the ellipsoidal modulation alone, as 
$\hat{q}_{\rm min} \sim 0.5$. This is so because the actual inclination of the system is substantially smaller than $90^{\circ}$.

\begin{figure} 
	\centering
	{  \includegraphics[height=5cm, width=8cm]{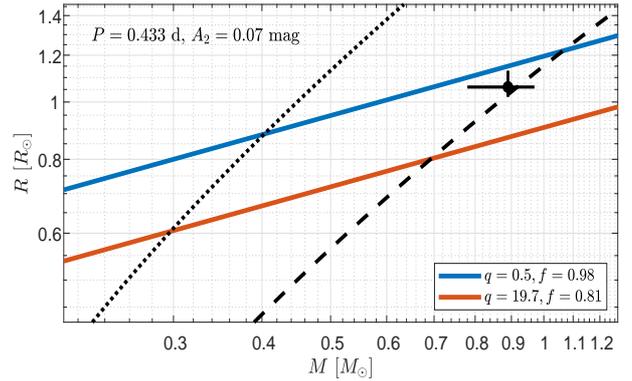}  }
	\caption{ Mass-radius relations for GRS 1124-683, derived by using $P = 0.433$ d, $A_2 = 0.07$ mag and $\alpha_2 = 1.2$, typical for the I band. The origin of each plotted line is described in the caption of Fig.~\ref{fig:M_R}. The black circle marks GRS 1124-683 position by adopting the values of \citet{wu16}. 
	}
	\label{fig:GRS1124}
\end{figure}

Our mass-radius relations are presented in Fig.~\ref{fig:GRS1124}, with the system position according to the values of \citet{wu16}, which is close to the line of fillout factor of unity, and close to the MS border of our permitted region.
%
%------------------------------------------------------------------
%------------------------------------------------------------------
%
\newpage
%================================
\section{Discussion}
\label{sec:discussion}
%================================

As shown above, one can derive the mMMR of a system, based on the ellipsoidal amplitude alone, assuming an inclination of $90^{\circ}$ and a fillout factor close to unity. The mMMR is substantially smaller than the actual mass ratio in most cases, and therefore an mMMR larger than unity can be used to identify ellipsoidal variables that might have a compact secondary.

Note that this approach cannot find all binaries with compact objects.  Systems with {\it dormant} BHs (of NSs) for which the primary did not evolve yet to fill its Roche lobe
%will be missed by our approach. 
%Therefore the ellipsoidal modulation necessarily 
do not display large enough amplitudes to yield mMMR larger than unity. In those cases, only additional information on the mass and radius of the primary can lead to a derivation of the MMR, which is always larger than the mMMR and is closer to the actual mass ratio. 

On the other hand, many BH and NS binaries with primaries that fill their Roche lobes are dormant, as they happen to be in an X-ray quiescent phase, in between outbursts \citep[e.g.,][]{cackett05, knevitt14}. For these systems, the mMMR could be significantly larger than unity. Ellipsoidal variables of this kind, favoured by our searching approach, are more likely to be detected.

The photometric search faces another problem --- the assumption that the inclination is close to $90^{\circ}$. Even systems with a fillout factor close to unity can yield small mMMR, if their inclination is low, as in the two cases discussed above, with actual mass ratios of $\sim 10$. This would cause many systems to be missed by a search based on the ellipsoidal amplitude alone.

This problem is inherent to spectroscopic and photometric searches alike. Fortunately, the random orientations of systems tend to favor inclinations of $90^{\circ}$, as the expected value for a sample of binaries is $<\sin i>=\pi/4=0.78$. 
In a practical search, one might consider lowering the threshold of $\hat{q}_{\rm min}$ that identifies BH candidates, allowing for binaries with $\sin i < 1$ and $f< 1$.
This is also true for the MMR, if the mass and radius of the primary are known. 
To determine the best threshold for a given project one can apply the False Discovery Rate (FDR) approach of \citet{benjamini95} that controls the purity of the resulting sample of discoveries.

Our method relies on identifying large sets of stellar light curves as ellipsoidal variables. 
For that we can use ellipsoidal-variables catalogs that are available for a few photometric surveys, such as CATALINA \citep{drake14}, ASAS-SN \citep{pojmanski02,pigulski09}, OGLE \citep{soszy16, pawlak16}, and {\it Kepler} \citep{kirk16}. For a detailed review of photometric-variables catalogs see the introduction of \cite{drake14}. 
In the future, the space mission Gaia, which follows the photometry of more than a billion stars \citep{Gaia2016}, is  expected to publish a large catalog of variable stars, as done for DR2 \citep{eyer19}. This will probably be the most extensive catalog of ellipsoidal variables.

For future photometric surveys, if a variable-star catalog is not available, one can identify the ellipsoidal variables by using, for example, classification methods similar to the ones described in \cite{eyer19}. Such methods can use a combination of the stellar position in the color-magnitude diagram, if available, together with Fourier-harmonics amplitude ratios \citep{soszy09}, to distinguish between ellipsoidals and other variables. 

Obviously, any catalog may include misclassified ellipsoidal variables, especially contact binaries and stellar rotational variables with stable periodicity. Those might result in false-positive compact-object candidates. Our false-positive ratio depends on the contamination of the ellipsoidal catalogs.

In the immediate next stage of this project, we are applying our technique to the large samples of short-period ellipsoidal variables \citep{soszy16, pawlak16} identified by the OGLE project \citep{udalski15}.

%The availability of large photometric surveys in the near future, the LSST \citep[e.g.,][]{abell09,ivezic19} in particular, increases the potential of the search proposed here. 

The photometric search we envision is focused on short-period binaries. The ellipsoidal amplitude falls as the period squared, so the search is sensitive up to at most $\sim5$-day binaries for MS stars. 
Note that for late-type MS stars the search is sensitive to NS companions as well, as the mass ratio can be around $3$ for a K-star of $\sim 0.5 M_{\odot}$ and an NS with a mass of $\sim 1.4 M_{\odot}$ (see a review by \citet[][]{enoto19}, and  \citet{haniewicz21}, for a detailed discussion of one specific system). 
A sample of BH/NS candidates should be followed by radial-velocity observations, to confirm they have massive companions. 
The availability of follow-up resources might, among others, determine the FDR threshold of a project.

Ellipsoidal searches can substantially enlarge the number of known short-period binaries with compact objects. This is  especially true because almost all known systems were discovered by their X-ray emission. We do not know yet of any short-period system with a compact object for which the optical star does not fill its Roche lobe.  The proposed project can reveal the systems for which the optical star is on its way to fill its Roche lobe, and therefore was not yet discovered. With the newly discovered systems, we will be able to better study the statistical characteristics of close systems with BHs or NSs companions.

%========================
\section*{Acknowledgments}
%========================
%
We are indebted to the referee, Andrew Cantrell, who contributed illuminating comments and suggestions on the previous version of the manuscript, helping us improve substantially the paper.
This research was supported by Grant No.~2016069 of the United States-Israel Binational Science Foundation (BSF) and by the Grant No.~I-1498-303.7/2019 of the
German-Israeli Foundation.
%
%========================
\section*{Data availability}
%========================
%
No new data were generated or analysed in support of this research.
%
%================================================
%

\bibliographystyle{mnras}
\bibliography{BHB_ellip_bib}
%
%%%%%%%%%%%%%%%%% APPENDICES %%%%%%%%%%%%%%%%%%%%%
%==============================================
%
%
%
\appendix
%====================
%
% Don't change 
% Don't change these lines
\bsp	% typesetting comment
\label{lastpage}
\end{document}